\documentclass[aps,pra,superscriptaddress,reprint, floatfix, longbibliography]{revtex4-1}
\usepackage{amstext,amssymb}
\usepackage{amsmath}
\usepackage{amsfonts}
\usepackage{xcolor}
\usepackage{graphicx}
\usepackage{bbold}
\usepackage[T1]{fontenc}
\usepackage{mathtools}
\usepackage[normalem]{ulem}
\definecolor{myred}{RGB}{228,26,28}
\definecolor{myorange}{RGB}{225,127,0}
\definecolor{mygreen}{RGB}{77,175,74}
\definecolor{mylila}{RGB}{152,78,163}
\def\Tr{\mbox{Tr}}

\newcommand{\jt}[1]{{\color{black}#1}}

\newcommand{\erg}{W_\mathrm{erg}(\varrho_N)} 
 
\newcommand{\ergc}{W_{\mathrm{erg}}^{(\mathrm{c})}(\varrho_{N})}
\newcommand{\erginc}{W_{\mathrm{erg}}^{(\mathrm{i})}(\varrho_{N})}

\begin{document}
\title{Charging quantum batteries via Otto machines: The influence of monitoring}
\author{Jeongrak Son}
\affiliation{Center for Theoretical Physics of Complex Systems, Institute for Basic Science (IBS), Daejeon 34126, Republic of Korea}
\affiliation{Department of Physics and Astronomy, Seoul National University, Seoul 08826, Republic of Korea}
\affiliation{School of Physical and Mathematical Sciences, Nanyang Technological University, 639673, Singapore}
\author{Peter Talkner}
\email[]{peter.talkner@physik.uni-augsburg.de}
\affiliation{Institut f\"{u}r Physik, Universit\"{a}t Augsburg, Universit\"{a}tsstra{\ss}e 1, D-86135 Augsburg, Germany} 
\author{Juzar Thingna}
\email[]{juzar\_thingna@uml.edu}
\affiliation{Center for Theoretical Physics of Complex Systems, Institute for Basic Science (IBS), Daejeon 34126, Republic of Korea}
\affiliation{Basic Science Program, Korea University of Science and Technology, Daejeon 34113, Republic of Korea}
\affiliation{Department of Physics and Applied Physics, University of Massachusetts, Lowell, MA 01854, USA}
\date{\today}

\begin{abstract}
The charging of a quantum battery by a four-stroke quantum machine that works either as an engine or a refrigerator is investigated. The presented analysis provides the energetic behavior of the combined system in terms of the heat and workflows of the machine, the average, and variance of the battery's energy as well as the coherent and incoherent parts of its ergotropy. To monitor the battery state its energy is measured either after the completion of any cycle or after a prescribed number of cycles is carried out. The resulting battery performances greatly differ for those two cases. During the first charging epoch with an engine, the regular measurements speed up the charging, whereas the gain of ergotropy is more pronounced in the absence of measurements. In a later stage, the engine fails to work as such while it still continues charging the battery that eventually reaches the maximally charged state in the absence of intermediate measurements and a suboptimally charged state for a periodically measured battery. For a refrigerator, the charging of the measured battery also proceeds faster during the first epoch. Only during the second stage when the machine fails to extract heat from the cold bath the influence of the measurements become less pronounced leading to rather similar asymptotic states for the two measurement scenarios.
\end{abstract}   
\maketitle
\section{Introduction}

The ongoing miniaturization of conventional devices also affects the problem of energy storage under the influence of quantum effects. Analogously to the classical case, a quantum battery (QB)~\cite{Campaioli2018} stores energy retrievable for later use with the aid of quantum mechanical effects such as entanglement or coherence for  efficient charging and extraction. 

Historically, a QB is an isolated quantum system undergoing unitary charging protocols. When an ensemble of such batteries is present, collective effects are responsible for  enhancing work extraction~\cite{Alicki2013, Hovhannisyan2013} or boosting the charging power~\cite{Binder2015, Campaioli2017} due to entanglement between the QBs. Perfectly isolating a quantum system is practically impossible, hence noisy QBs are studied to either test the impact of dissipation on the battery performance~\cite{Carrega2020,  Zakavati2021}, or to utilize noise as a valuable resource~\cite{Barra2019, Hovhannisyan2020, Tacchino2020}. There are also proposals to engineer decoherence-free subspaces that protect a quantum battery against dissipation~\cite{Dvira19, Munro20, Thingna21}. Moreover, interesting non-classical charging behaviors are observed if the charger is also explicitly modeled as another quantum system~\cite{Andolina2018, Farina2019, Andolina2019, Seah2021}, allowing for feedback-control~\cite{Mitchison2021}.

In this work, we explore the charging of a QB with $M$ equidistant energy levels by means of a two-level quantum Otto machine~\cite{Feldmann2003}. Similar two-level quantum chargers that are subjected to a drive have been explored before~\cite{Mitchison2021, Seah2021}, but never has the charger itself been considered as a thermal machine. The quantum machine performs its usual thermodynamic tasks as an engine, outputting work,  or as a refrigerator, cooling the cold bath, while additionally charging the battery. The scenario is similar to a fossil-fuel-based car engine that not only moves the car, but also charges up the battery that powers the peripheral devices. 

A similar set-up was studied to analyze the role of the number of cycles on the Otto engine's performance~\cite{Watanabe2017}. Watanabe et al. envisaged a $M$-level flywheel as an apparatus to which the engine performs work. For this device it was demonstrated that the average energy stored in the flywheel is not strictly proportional to the number of engine cycles when the coherence of the flywheel is retained, rather than being suppressed by projective energy measurements at the end of each cycle. The effect of  the suppression of coherence by repeated measurements on the performance of an Otto engine over large times was investigated in Ref.~\cite{Son2021}.

Inspired by Refs.~\cite{Watanabe2017, Son2021}, we study the complete energetics of a QB charged by a quantum Otto machine~\footnote{We classify Otto machines as {\it engines} if they perform work on an external agent or as {\it refrigerators} if they pump heat from a cold to a hot reservoir by a sequence of four strokes as sketched in Fig.~\ref{fig1:schematic}.}    over a given number of cycles including the ergotropy that quantifies the amount of work that can be extracted from the QB via unitaries~\cite{Pusz1978,Allahverdyan2004}. We demonstrate that frequent monitoring of the battery on the one hand increases the rate of internal energy charging of the battery but on the other hand leads to a lower ergotropy as compared to the QB that is not monitored. Apart from the battery's  energy we analyze the cycle-wise evolution of different work flows of the machine to the external control field and to the battery as well as of the total energy. In particular, we find  that after a certain number of cycles the machine ceases to work, whether it is supposed to perform as an engine supplying energy to the external control field or as a refrigerator cooling the hot bath. Despite the machine's breakdown, the QB continues being charged, albeit at a slower rate. Thus, in addition to the asymptotic behavior of the models under two different monitoring strategies, the amount of charging before the machine fails is introduced and examined as a new figure of merit.

The paper is organized as follows: Sec.~\ref{sec:setting} introduces the basic setup and defines the various engine and QB metrics studied in this work. Section~\ref{sec:charging} focuses on the main results of this paper, wherein we charge the QB either using an engine or a refrigerator. Finally, we conclude in Sec.~\ref{sec:conclusion}. In the Append.~\ref{appendix:engine_only} the working modes of an Otto machine are reviewed, in the Append.~\ref{appendix:switch_onoff} the energetic effect of periodically turning the battery-machine coupling on and off is estimated, \jt{and in Append.~\ref{appendix:energetics} the energetic cost of measuring energies of isolated systems is investigated.}   

\section{Setting up the stage}\label{sec:setting}
\subsection{Charging Protocol}\label{subsec:charging_setting}
The working fluid of the Otto machine, which is used to charge the battery, is taken as a two level system undergoing two work and two thermalization strokes within a machine cycle. During the work strokes the machine is coupled to an $M$-level QB. To avoid any, even indirect, contact of the QB with the heat baths during the thermalization strokes the QB is disconnected from the machine during these strokes. Thus, unlike standard Otto cycles, the working fluid of the machine, running in either mode as an engine or a refrigerator, is not only exchanging work with the external drive, but, as we shall demonstrate, may also deliver energy to the QB and hence charge it.    

During the work strokes the combined machine-battery system is described by the following Hamiltonian  
\begin{equation}
    H(t) = H_{\mathrm{M}}(t) + H_{\mathrm{B}} + H_{\mathrm{MB}}(t), 
\end{equation}
where the Hamiltonian of the working fluid $ H_{\mathrm{M}}(t)$ is given by
\begin{eqnarray}
    H_{\mathrm{M}}(t) &=& \Delta\sigma_{x} + \xi(t)\sigma_{z},\label{eq:H_engine}
\end{eqnarray}    
with the Pauli matrices $\sigma_i$, $i=x,y,z$, a constant tunnel parameter $\Delta$ and a time-dependent field $\xi(t)$ that is specified below. The battery is modeled as an $M$-level energy reservoir with an equi\-spaced spectrum accordingly described by the Hamiltonian 
\begin{eqnarray}    
    H_{\mathrm{B}} &=& \omega\sum_{l=0}^{M-1}l\: \Pi_l,
\label{eq:H_battery}
\end{eqnarray} 
with the level-spacing $\omega >0$, and the projection operators $\Pi_l = \lvert l\rangle\langle l\rvert$, $l=0 \ldots M-1$  projecting onto the $l$th battery levels. The interaction of the machine and the QB is governed by
\begin{eqnarray}
    H_{\mathrm{MB}}(t) &=& g(t)\sigma_{x}\otimes q.
\label{eq:H_interaction}
\end{eqnarray}
where $q =\sum_{l=0}^{M-2}\sqrt{l+1}(\lvert l\rangle\langle l+1\rvert + \rm{h.c.}) $ denotes the position operator of the QB. The coupling coefficient $g(t)$ is specified below. Throughout, we set $\hbar = k_{\mathrm{B}} = 1$.

\begin{figure}[t!]
    \centering
   \includegraphics[width = \columnwidth]{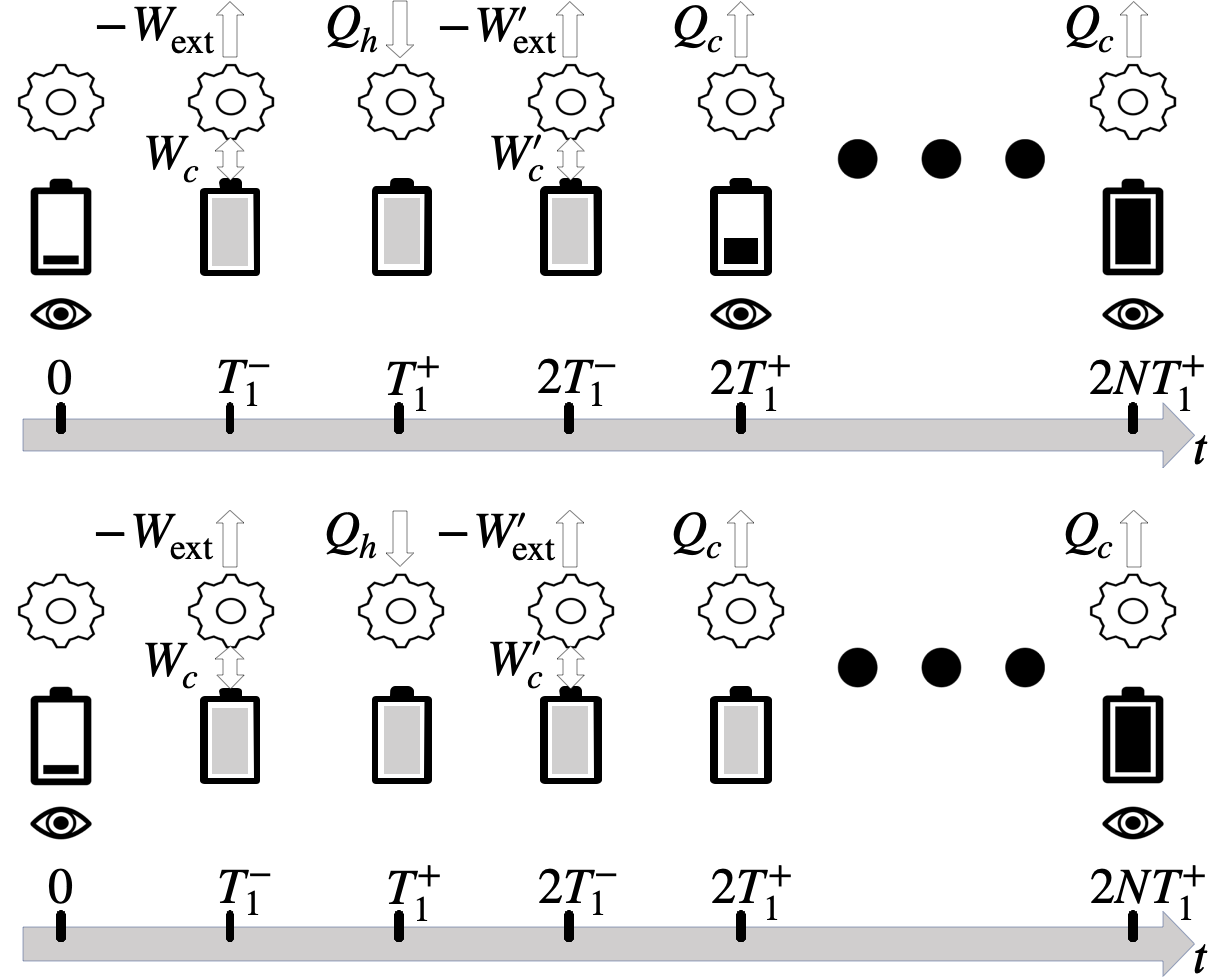}
    \caption{An illustration of a quantum Otto machine charging a QB. The working fluid (illustrated as a gear) being initially prepared in a thermal equilibrium state at the low inverse temperature $\beta_c$ undergoes a compression (stroke 1) between $0 \leq t \leq T_1^-$ while staying in contact with the battery but otherwise in isolation, followed by perfect thermalization by a contact with a hot bath at the inverse temperature $\beta_h$ (stroke 2) between $T_1^- < t < T_1^+$. During the subsequent third stroke the working fluid undergoes, again in contact with the battery but else in isolation, the time-reversed process (expansion) of the first stroke between $T_1^+ \leq t \leq 2T_1^-$. In the final stroke (4) the contact with a cold heat bath between $2T_1^- < t < 2T_1^+$ leads to perfect thermalization at the initial cold inverse temperature $\beta_c$. During the two heat strokes the battery is isolated from the machine. The heat-strokes (stroke 2 and 4) are assumed to be instantaneous. During the work-strokes, the machine does work on the external field ($-W_{\rm ext}$, $-W'_{\rm ext}$) and exchanges energy ($W_{\rm c}$, $W'_{\rm c}$) with the QB. The top row illustrates a diagnostic scheme in which the battery is measured at the end of each cycle (illustrated by an eye symbol), whereas the bottom row shows a battery that is measured only at the end of $N$ cycles.}
    \label{fig1:schematic}
\end{figure}
The working fluid is subjected to four separate strokes per cycle. It first undergoes a compression of duration $T_1$, in which the control parameter $\xi(t)= vt$ ($0 \leq t \leq T_1$) is varied linearly such that the energy gap in the two-level working fluid increases. The compression stroke is carried out in thermal isolation, with the QB connected to the machine ($g(t) = g$). Thus, the machine-battery state evolves with the operator
\begin{equation}\label{eq:evop}
    U = \mathcal{T}\exp\left[-i \int_{0}^{T_1}\,dt\,H(t)\right].
\end{equation}
where $\mathcal{T}$ denotes time-ordering. During this stroke, the machine performs work on the external field in the engine mode and absorbs energy in the refrigerator mode. At the end of the compression-stroke, the working fluid is connected to a hot bath and $\xi(t) = vT_1$ is no longer varied. 

As mentioned, during the heat-strokes, the machine-battery coupling is turned off, i.e., $g(t) = 0$ to avoid thermalizing the QB. During  the hot heat stroke of the duration $T_2$ we assume that the working fluid reaches the inverse target temperature $\beta_h$. As a result of the hot heat-stroke, the joint machine battery state is given by the operation $\Phi^h$, acting on the respective joint density matrix $\rho$ before this stroke started such that,
\begin{eqnarray}\label{eq:Gibbsmap}
    \Phi^{h}(\rho) &=& \tau_{h} \otimes U_B \Tr_{\mathrm{M}}[\rho] U^\dagger_B,
\label{eq:therm_engine}\\
    \tau_{h} &=& \frac{\exp[-\beta_{h}H_{\mathrm{M}}(T_1)]}{\Tr{\left[\exp[-\beta_{h}H_{\mathrm{M}}(T_1)]\right]}},\label{eq:Gibbs_state}
\end{eqnarray}
where $\Tr_{\text{M}}$ and $\Tr= \Tr_\mathrm{M}\Tr_\mathrm{B}$ denote the traces over the working fluid and the combined system, respectively, with the trace over the battery Hilbert space marked by $\Tr_\mathrm{B}$. Further, $\tau_{h}$ indicates the canonical Gibbs state of the hot working fluid; moreover,  $U_\mathrm{B} =e^{-i H_\mathrm{B} T_2}$ renders the free time-evolution of the battery during the thermalization of the working fluid. For the sake of simplicity, we assume that $U_B = \mathbb{1}$. This may be achieved by an instantaneous thermalization of the working fluid or by choosing the thermalization time as an integer number $n$ of the principal period of the battery, $T_2= 2 \pi n /\omega$. The deteriorating influence of $T_2$ for non-integer  multiples of the principal battery period on the charging of the battery is shortly discussed below. The effect of switching the battery-machine interaction periodically on and off on the battery energy is difficult to specify for the full system. \jt{In Append.~\ref{appendix:switch_onoff}, for a simplified model, }the energetic impact of the on-off switching on the battery is found to be negligible for any value of $T_2$.   

After the heat-stroke the working fluid is subjected to an expansion-stroke, which follows the time-reversed protocol of the compression such that $\xi(t) = v(2T_1+T_2 - t)$ for $T_1 +T_2 \leq t \leq 2T_1 +T_2$.  The states are evolved with $\tilde{U} = CU^{\dagger}C$, where $U$ is given by Eq.~(\ref{eq:evop}) and $C$ is the complex conjugation operator. This ensures that the energy spectrum of the working fluid undergoes a cyclic variation. The expansion is followed by a heat-stroke at a cold temperature. As a result, the density matrix of the combined system  is described by the action of an  operation $\Phi^c$  applied to the density matrix at the time $2T_1+T_2$. The operation $\Phi^c$ is given by Eq.~(\ref{eq:Gibbsmap}) with the hot inverse temperature $\beta_h$ replaced by the cold inverse temperature $\beta_c$ and the Hamiltonian $H_\mathrm{M}(T_1)$  being replaced  by $H_\mathrm{M}(0)$. When the machine operates as an engine, it supplies work to both the battery and the external field, whereas in the refrigerator mode it absorbs work from the external field cooling down the cold bath and supplying work to the QB. Therefore, the operation describing the map of the joint density matrix caused by a complete cycle becomes
\begin{equation}
    G(\rho) = \Phi^{c}(\tilde{U}\Phi^{h}(U\rho U^{\dagger})\tilde{U}^{\dagger}).
    \label{cycle_tot}
\end{equation}
The above operation is then repeated $N$-times to obtain the machine-battery density matrix after $N$ cycles.

In order to monitor the charging of the battery we consider the same two  diagnostic schemes as described in Ref.~\cite{Watanabe2017}. According to one scheme, the energy of the battery is projectively measured after the completion of every cycle. In this way any coherence in the energy eigenbasis of the battery that might have built up during a cycle is erased. Thus, the machine-battery density matrix $\tilde{\rho}_{N+1}$ for a repeatedly measured battery at the end of $(N+1)$th cycle becomes
\begin{eqnarray}
    \tilde{\rho}_{N+1} &=& \sum_l \Pi_l G(\tilde{\rho}_{N}) \Pi_l
    \equiv \tilde{G}(\tilde{\rho}_{N}),
\end{eqnarray}
with the projection operators $\Pi_l$ as defined in Eq,~(\ref{eq:H_battery}). In the other scheme only a single projective battery energy measurement is performed after a prescribed number of $N_f$ cycles,   
\begin{equation}
    \rho_{N+1} = G(\rho_{N})
\end{equation}
for $N+1<N_f$. Eventually, after $N_f$ cycles the battery energy is projectively measured. \jt{In either case, such a measurement is performed immediately after each contact with the cold heat bath when the battery is disconnected from the machine. This ensures that there is no additional energetic cost due to measurements as shown in Append.~\ref{appendix:energetics}.} However, the extraction of energy from the battery by means of a unitary process might advantageously be performed before any energy measurement of the battery to make use of energy stored in coherence (see Sec.~\ref{subsec:b_metric}). Figure~\ref{fig1:schematic} provides a sketch of the modes of operation of the battery-machine device under the two  monitoring regimes. For the sake of simplicity, in the following the thermalization strokes are assumed to work instantly, i.e., $T_2 \equiv T_1^+ - T_1^- \rightarrow 0$, if not explicitly stated otherwise. Throughout this work, all quantities corresponding to a repeatedly measured battery are distinguished by a tilde.

\subsection{Machine Metrics}\label{subsec:m_metric}
In order to get a complete energetic picture of the charging process, projective energy measurements of the thermally isolated working fluid are performed at the beginning and the end of each work stroke  as described in more detail in Refs.~\cite{Ding2018,Son2021}. Due to the here assumed perfect thermalization of the heat strokes these measurements do not influence the dynamics of the engine~\cite{Son2021}. Therefore the average heats exchanged with the hot and cold reservoirs during the $(N+1)$th cycle are given by the differences of the average working fluid  energies after and before the respective strokes and hence result as: 
\begin{eqnarray}
    \langle Q^{h}_{N+1}\rangle &=& \Tr_\mathrm{M}\left[H_{\mathrm{M}}(T_1)\left(\tau_{\mathrm{h}} - \Tr_{\mathrm{B}}[\rho_N^{T_1^-}]\right)\right],\\
    \langle Q^{c}_{N+1}\rangle &=& \Tr_\mathrm{M}\left[H_{\mathrm{M}}(2T_1)\left(\tau_{\mathrm{c}} - \Tr_{\mathrm{B}}[\rho^{2T_1^-}_{N}] \right)\right],\label{eq:heat_def}\nonumber
\end{eqnarray}
Above, the machine-battery density matrix after the compression stroke of the $(N+1)$th cycle is given by $\rho^{T_1^-}_N = U\rho_{N}U^{\dagger}$ and the one after the expansion stroke becomes, $\rho^{2T_1^-}_{N} = \tilde{U}\Phi^{h}\left(U\rho_{N}U^{\dagger}\right)\tilde{U}^{\dagger}$. 

Likewise, the total average work done on the machine-battery system is evaluated as the change of internal energy of the composite before and after the work-strokes, giving
\begin{eqnarray}
    \langle W_{N+1}\rangle &=& \Tr\left[H(2T_1) \, \rho_N^{2T_1^-} - H(T_1)\, \rho_N^{T_1^+}\right]\nonumber\\
    &+& \Tr\left[H(T_1)\rho_N^{T_1^-} - H(0)\rho_N \right],\label{eq:work_def}
\end{eqnarray}
where $\rho_{N}^{\rm T}$ denotes the density matrix at the time $T$ after the beginning of the $(N+1)$th cycle. Specifically, the time $T_1^-$ refers to the final time at the end of the compression stroke, $T_1^+$ to the time at the end of the hot heat-stroke, $2T_1^-$ the end of the expansion stroke, and $2T_1^+$ marks the end of the cold heat-stroke.  The above expressions for the heats and the total work are valid for a device whose battery is monitored only after the completion of a prescribed number of cycles, $N_f$. For the alternative periodically monitoring scenario the respective averages are obtained by replacing the state $\rho$ by the monitored state $\tilde{\rho}$. Finally we note that at all above measurement times $T^{\pm}_1$, $2 T^{\pm}_1$ the coupling between machine and battery vanishes, $g(t)=0$, such that the energies of the working fluid and of the battery can be determined simultaneously \jt{without an extra energetic cost (see Append.~\ref{appendix:energetics})}.   

While, due to the perfectly thermalizing heat strokes, the machine's reduced state evolves cyclically if it is initiated in the cold Gibbs state, the machine-battery system undergoes a transient dynamics during which the battery is charged until it eventually approaches a periodic state.   

\subsection{Battery Metrics}\label{subsec:b_metric}
The QB is initialized in its ground state  at zero energy; hence, the average internal energy stored in the QB after $N$ cycles is given by,
\begin{equation}
E_{N} = \langle H_{\rm B}\rangle = \Tr{[H_{\rm B}\rho_N]}.
\label{eq:EN_def}
\end{equation}
A related, also vital, performance figure of a QB is the variance of its energy,
\begin{equation}
   \sigma_E^2(N) =  \langle H_{\rm B}^{2}\rangle - \langle H_{\rm B}\rangle^2 = \Tr{[H_{\rm B}^2\rho_N]} - \Tr{[H_{\rm B}\rho_N]}^2.
\label{DE}
\end{equation}
The variance quantifies the fluctuations of the battery energy after $N$ cycles.  
As a result of the periodicity of the machine state the energy of the working fluid does not change after a cycle. Hence,  the average heats and total work per cycle are related to the energy increase of the battery by the relation 
 \begin{equation}
    \langle Q^{h}_{N}\rangle + \langle Q^{c}_{N}\rangle + \langle W_{N}\rangle = E_{N} - E_{N-1}. \label{eq:noncyclic}
\end{equation}
 \jt{In passing, we note that for an engine this equation relates the sum of the engine efficiency $\eta^e_N=-\langle W_N \rangle/\langle Q^h_N\rangle$ and the the charging efficiency $\eta^c_N = (E_N - E_{N-1})/\langle Q^h_N\rangle$ to the ratio of heats exchanged with the cold and the hot heat baths, $r_N= \langle Q^c_N \rangle/  \langle Q^h_N \rangle$ according to $\eta^e_N + \eta^c_N =1 +r_N$ with $-1<r_N<0$ for an engine.}
 
In general, it is impossible to extract all the stored internal energy $E_N$ of the QB~\cite{Allahverdyan2004, Niedenzu2019}. The maximal amount of energy that can be extracted via unitary processes from a QB after $N$ charging cycles is quantified by the ergotropy that, according to Ref.~\cite{Allahverdyan2004}, can be expressed as  
\begin{eqnarray}
    W_{\mathrm{erg}}(\varrho_N) &=& \max_{U}\left[\Tr[\varrho_N H_\mathrm{B}] - \Tr[U\varrho_{N}U^{\dagger}H_\mathrm{B}]\right] \nonumber \\
    &=&\sum_{i,j}\epsilon_{i}r_{j}(\lvert\langle\epsilon_i|r_j\rangle\rvert^{2} - \delta_{ij}), 
\label{erg}
\end{eqnarray}
where the eigenvalues $r_j$ of the battery density matrix are labeled such that $r_j$  decreases  with increasing index $j$ while the eigenvalues of the battery Hamiltonian, $\epsilon_j= \omega j$, increase. For a regularly measured battery, the according ergotropy assumes the same form as in Eq.~(\ref{erg}) with $\varrho_N$ replaced by $\tilde{\varrho}_N = \Tr_{\mathrm{M}}[\tilde{\rho}_N]$ as well as $r_j$ replaced by the eigenvalues $\tilde{r}_j$ of $\tilde{\varrho}_N$. Furthermore, the ergotropy can be divided into coherent and incoherent contributions~\cite{Francica2020} that quantify whether the extractable energy is stored in a superposition of the quantum states (coherence) or via a population inversion like in a laser. The incoherent contribution
\begin{equation}
\label{eq:Winc}
	W_{\mathrm{erg}}^{\mathrm{(i)}}(\varrho) = W_{\mathrm{erg}}\left(\sum_{i}\Pi_i\varrho\Pi_i\right)
\end{equation}
equals the ergotropy of the decohered state. In other words, if we measure the battery energy at the end of $N$ cycles, thereby killing all coherence, $W_{\mathrm{erg}}^{(i)}(\varrho_N)$ would be the remaining extractable energy after the final measurement. The coherent contribution then naturally emerges as,
\begin{equation}
\label{eq:Wc}
    W_{\mathrm{erg}}^{\mathrm{(c)}}(\varrho) = W_{\mathrm{erg}}(\varrho) - W_{\mathrm{erg}}^{\mathrm{(i)}}(\varrho).
\end{equation}

Finally we recall that the ergotropy presents an upper bound on the work. If only a restricted class of unitary operations is available to the experimenter, this bound can be inaccessible.  

\section{Effective charging of the battery}\label{sec:charging}
In this section, we design protocols to charge a QB using a thermal machine. First we note that for an Otto machine with a two-level working fluid and with perfectly thermalizing heat strokes the machine parameters can be chosen such that, in the absence of a battery, it either works as an engine, or as a refrigerator. To work as an engine, as a necessary condition \jt{$\beta_h \epsilon_h < \beta_c \epsilon_c$} must hold. Further a parameter $\alpha$ characterizing the rapidity of the work-strokes must not exceed a limiting value depending on the compression factor \jt{$\epsilon_h/\epsilon_c$}. For a refrigerator \jt{$\beta_h \epsilon_h > \beta_c \epsilon_c$} must hold and $\alpha$ is bounded by a function of \jt{$\beta_h \epsilon_h$} and \jt{$\beta_c \epsilon_c$} which is independent of the compression factor. For more details \jt{on the functioning of an isolated Otto engine with a two-level system as working substance see Append.~\ref{appendix:engine_only}}.

In the presence of a battery, which is charged by the machine, an analytic solution is no longer available. We therefore present results of a numerical study, first of machines that in isolation would work as engines, and then shortly consider the case of refrigerators.    

\subsection{Initializing the Machine as an Engine}\label{subsec:engine}
\begin{figure}[t!]
    \centering
    \includegraphics[width=\columnwidth]{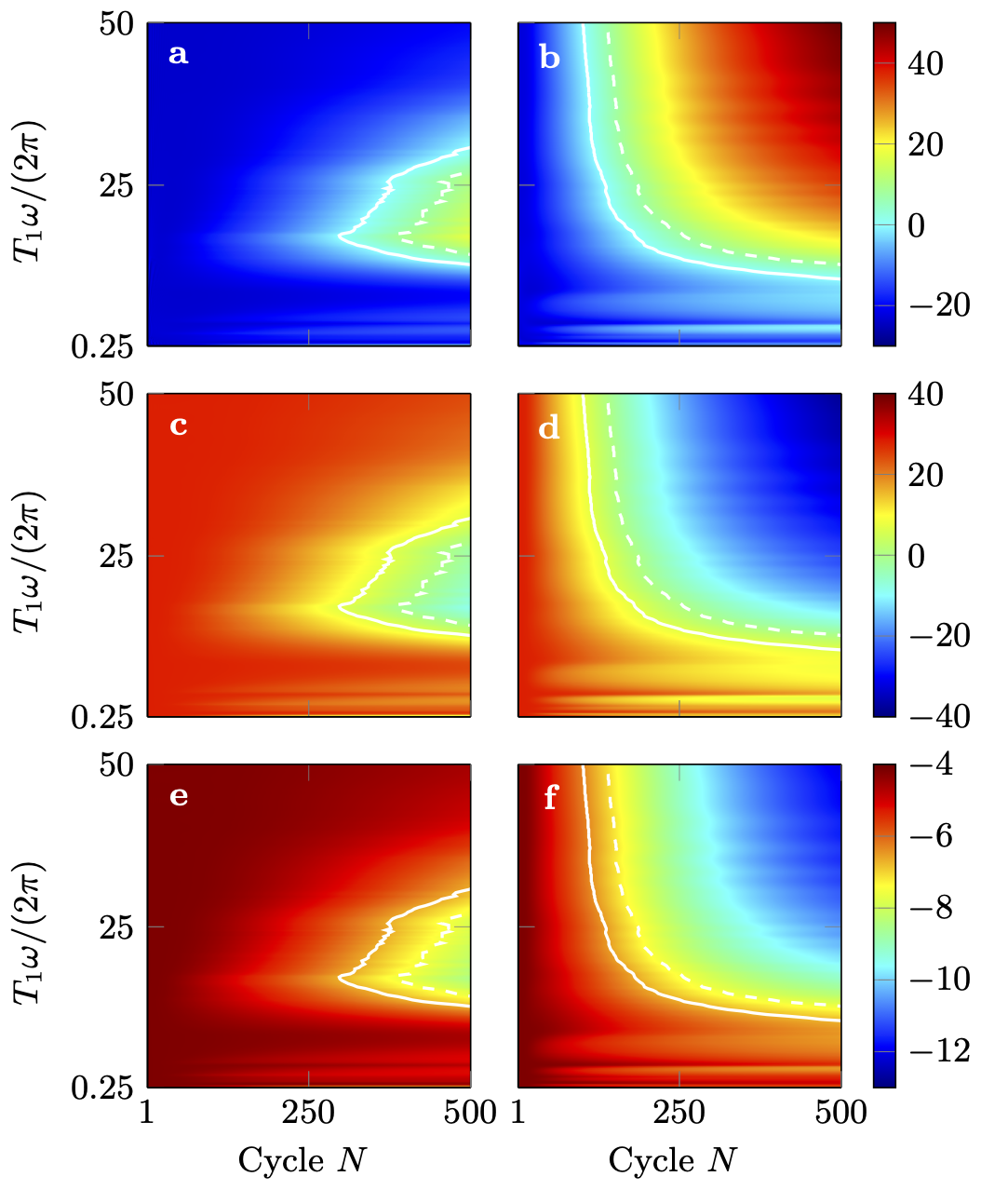}
    \caption{Average work $\langle W_N\rangle$ ({\bf a} and {\bf b}), average heat from the hot bath $\langle Q^h_N\rangle$ ({\bf c} and {\bf d}), and average heat from the cold bath $\langle Q^c_N\rangle$ ({\bf e} and {\bf f}) as functions of the cycle number $N$ and the period of the work-stroke $T_1$. Positive values (on the color scale) indicate the inflow of quantities to the working fluid, and negative indicate the outflow. The left column ({\bf a}, {\bf c}, and {\bf e}) corresponds to the unmeasured battery $\langle W_N \rangle$ and $\langle Q^{h(c)}_N \rangle$, whereas the right column ({\bf b}, {\bf d}, and {\bf f}) describes the periodically measured ($\langle \tilde{W}_N \rangle$ and $\langle \tilde{Q}^{h(c)}_N \rangle$) battery. The white solid lines mark the transition from a properly working engine to a failing engine that absorbs heat from the hot bath, emits to the cold bath, and absorbs work from the classical drive. The white dashed lines denote the transition to a situation in which the working fluid emits heat to both baths while absorbing work. Other parameters are set to $\epsilon_c=\Delta = 30\omega$, $v T_1 = 200\omega$, $g = \omega$, $\beta_h^{-1} = 200\omega$, $\beta_c^{-1} = 20\omega$, and QB levels $M=300$.}
    \label{fig:eng_WQ}
\end{figure}

\begin{figure}[t!]
    \centering
    \includegraphics[width=\columnwidth]{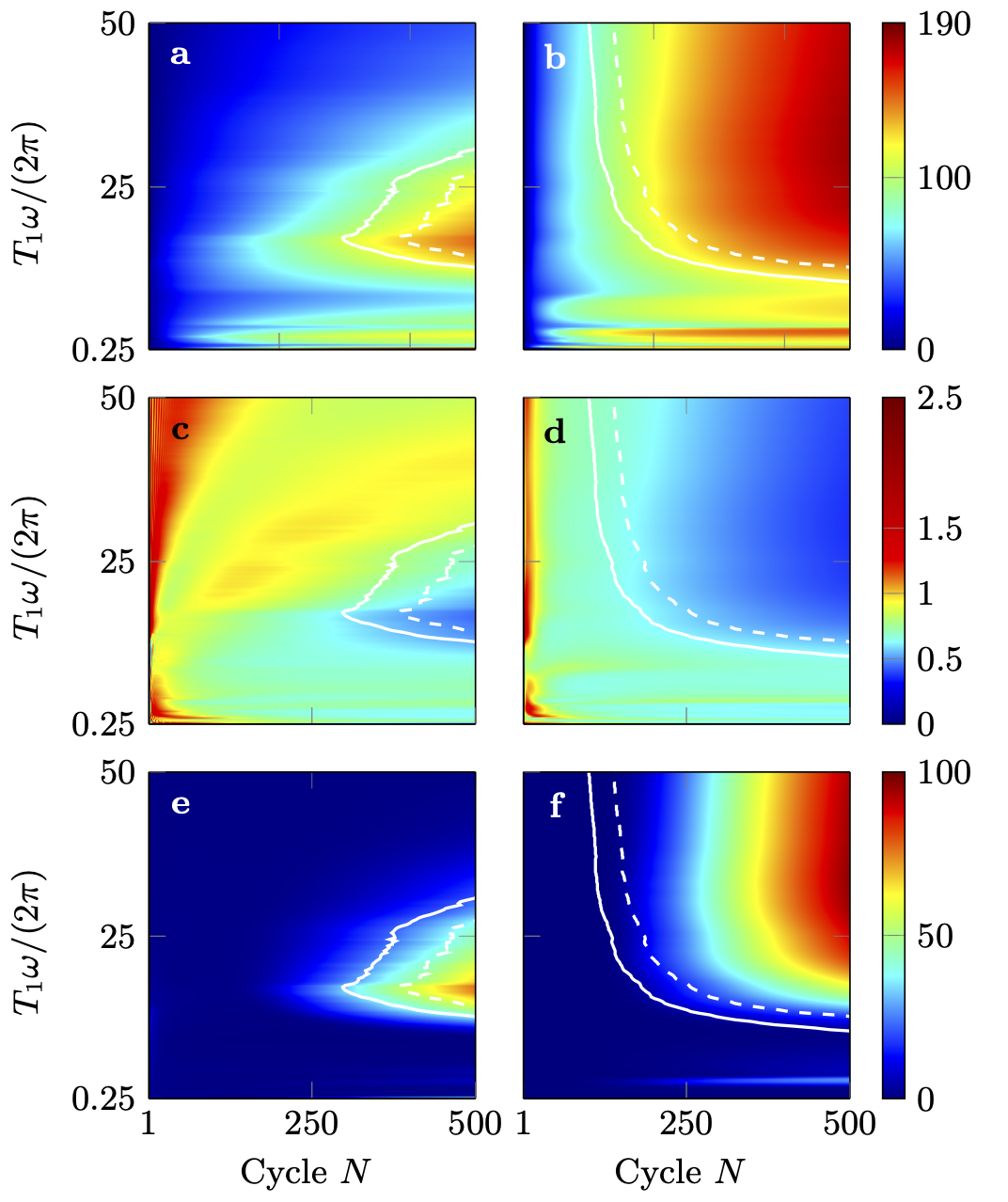}
    \caption{Average internal energy $E_N$ ({\bf a} and {\bf b}), coefficient of variation of the internal energy $C_N = \sigma_E(N)/E_N$ ({\bf c} and {\bf d}), and ergotropy $W_{\mathrm{erg}}(\varrho_N)$ ({\bf e} and {\bf f}) for a QB charged by an Otto engine. The left column ({\bf a}, {\bf c}, and {\bf e}) represents the results for an unmeasured battery, whereas the right column ({\bf b}, {\bf d}, and {\bf f}) is for the measured QB. The battery metrics are from the same parameters used to characterize the engine in Fig.~\ref{fig:eng_WQ}. As there, white solid and dashed lines correspond to the transition between a working engine to a failing engine emitting heat to the cold, and to both, the cold and the hot baths, respectively. Note here that the energy fluctuations of the QB, except for low numbers of cycles wherein the battery undergoes a rapid charging, are of the same order of magnitude as that of the internal energy stored in the QB.} 
    \label{fig:eng_E_ergo}
\end{figure}

\begin{figure}[t!]
    \centering
    \includegraphics[width=\columnwidth]{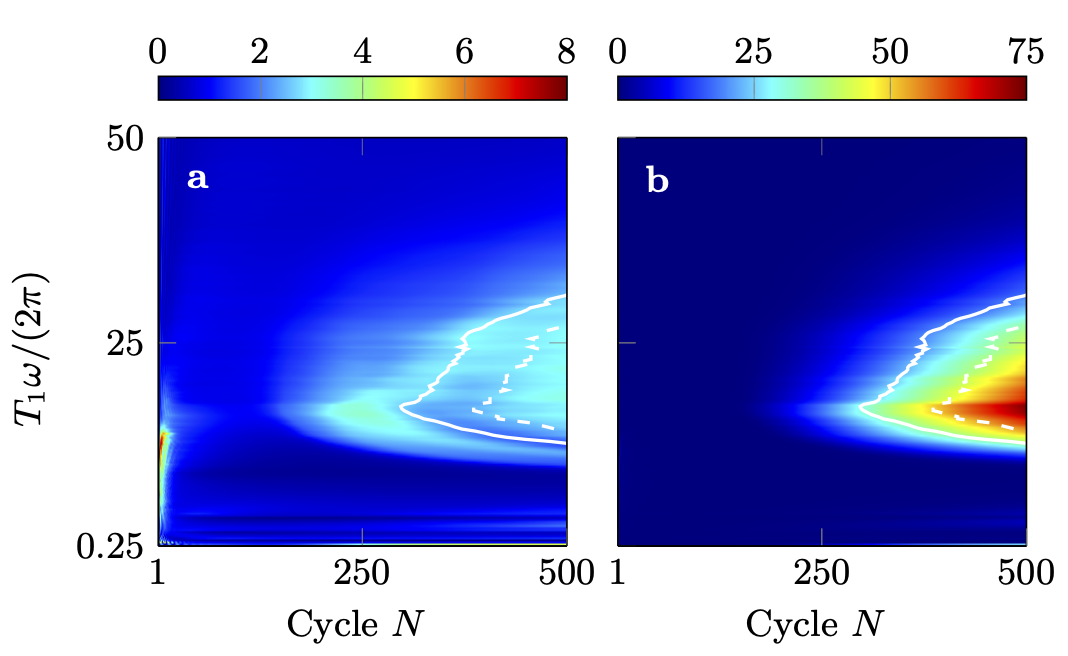}
    \caption{Coherent ergotropy $\ergc$ (panel {\bf a}) and incoherent ergotropy $\erginc$ (panel {\bf b}) for an unmeasured QB. The total ergotropy $W_{\mathrm{erg}}(\varrho_{N})$ is plotted in Fig.~\ref{fig:eng_E_ergo}{\bf c}. The parameters are identical to the ones used in Figs.~\ref{fig:eng_WQ} and \ref{fig:eng_E_ergo}. Note that the color scale for {\bf a} is one order of magnitude smaller than that of {\bf b}.}
    \label{fig:eng_ergo_coh_incoh}
\end{figure}

When the machine starts in the above specified parameter regime of a working engine  and the machine-battery interaction is not too strong, the machine  continues to perform as an engine for a number of cycles $N^*$ after which the average work changes its sign as shown in Fig.~\ref{fig:eng_WQ}. \jt{Nevertheless, the heat exchanged with the hot bath} remains positive until it turns negative after $N^{\#}$ cycles and finally approaches an asymptotic value. These critical cycle numbers depend on the strength of the engine-battery coupling $g$, which for all results presented below has the same value $g=\omega$. For larger values of $g$ the critical cycle numbers rapidly decrease until at a characteristic coupling strength the engine immediately stops working when coupled to the battery. The values of $N^*$ and $N^{\#}$ in dependence of the duration $T_1$ of the work stroke are indicated in the Figs.~\ref{fig:eng_WQ}--\ref{fig:eng_ergo_coh_incoh} as white solid and white dashed lines, respectively.

For both measurement protocols the dependence of the two average heats $\langle Q^h \rangle$ and $\langle Q^c \rangle$ on the number of cycles and on the duration of the work strokes is synchronous with the respective dependence of the total average work  in the sense that along any line of constant average work the heats stay also almost constant.  For example, in the absence of battery energy measurements, along the line with $\langle W_{N^*} \rangle = 0$  the two average heats take the approximate values  $\langle Q^h_{N^*}\rangle \simeq 7 \omega$ and $\langle Q^c_{N^*}\rangle\simeq -7 \omega$, independent of $T_1$. 

In the presence of periodic battery energy measurements, the average work as well as the average heats behave quite differently compared to the unmeasured battery scenario (see Fig.~\ref{fig:eng_WQ}). In particular, the engine reaches the number of cycles, where it starts to fail, earlier than the unmeasured battery, i.e., at any given $T_1$ one finds $\tilde{N}^* < N^*$. Further the number of cycles until the machine fails to work as an engine decreases with increasing duration of the work strokes. The fact that the unmeasured battery leads to a later failure of the engine might have its reason in the persistence of the battery coherence over several cycles, which is destroyed each time when the battery is measured. Also the contour line with $\langle \tilde{W}_{N^*} \rangle =0$ differs significantly from its unmeasured appearance, yet  $\langle \tilde{Q}^h_{N^*} \rangle$ and $\langle \tilde{Q}^c_{N^*} \rangle$ are almost constant along this curve at $\approx \pm 7 \omega$, respectively.  

In order to characterize the battery during the charging process, the average battery energy, its ergotropy, and the coefficient of energy variation as defined  by the Eqs.~(\ref{eq:EN_def}),~(\ref{DE}), and~(\ref{erg}), respectively, are displayed in Fig.~\ref{fig:eng_E_ergo} for the two considered monitoring scenarios. The average battery energy after $N$ cycles is the sum of the energies transferred by the engine to the battery in the previous cycles. At any fixed work stroke duration $T_1$ it monotonically increases with the number of cycles indicating that the engine moves energy to the battery not only in the regime of a properly working engine when $N < N^*$ but also in the two regimes of no work output while heat is still absorbed from the hot bath, i.e., for $N^* < N < N^{\#}$ and also when finally heat is emitted also to the hot bath for $N >N^{\#}$.         

The variation of the battery energy $\sigma_E(N)/E_N$ is depicted in the middle row of Fig.~\ref{fig:eng_E_ergo}; it  varies roughly around 1 being  largest at small $N$ with a tendency to decrease with $N$. Overall the periodically measured battery exhibits a lower variability and hence appears more reliable than an unmeasured one.   

The periodically measured battery approaches its asymptotic energy content much faster than the unmeasured one. Likewise, the asymptotic value of the ergotropy is approached after a substantially smaller number of cycles  for a periodically measured battery. Due to the measurement-induced elimination of coherence, the coherent part of the ergotropy, as defined in Eq.~(\ref{eq:Wc}), vanishes. Yet, even for a completely unmeasured battery the coherent part turns out to be substantially smaller than the incoherent one, see Fig.~\ref{fig:eng_ergo_coh_incoh}, and therefore also the ergotropy of an unmeasured battery mainly results from a partly inverted population of its energy eigenstates rather than from coherence. A final battery energy measurement of an otherwise unmeasured battery has consequently no tangible effect on the ergotropy. 

\begin{figure}[t!]
    \centering
    \includegraphics[width=\columnwidth]{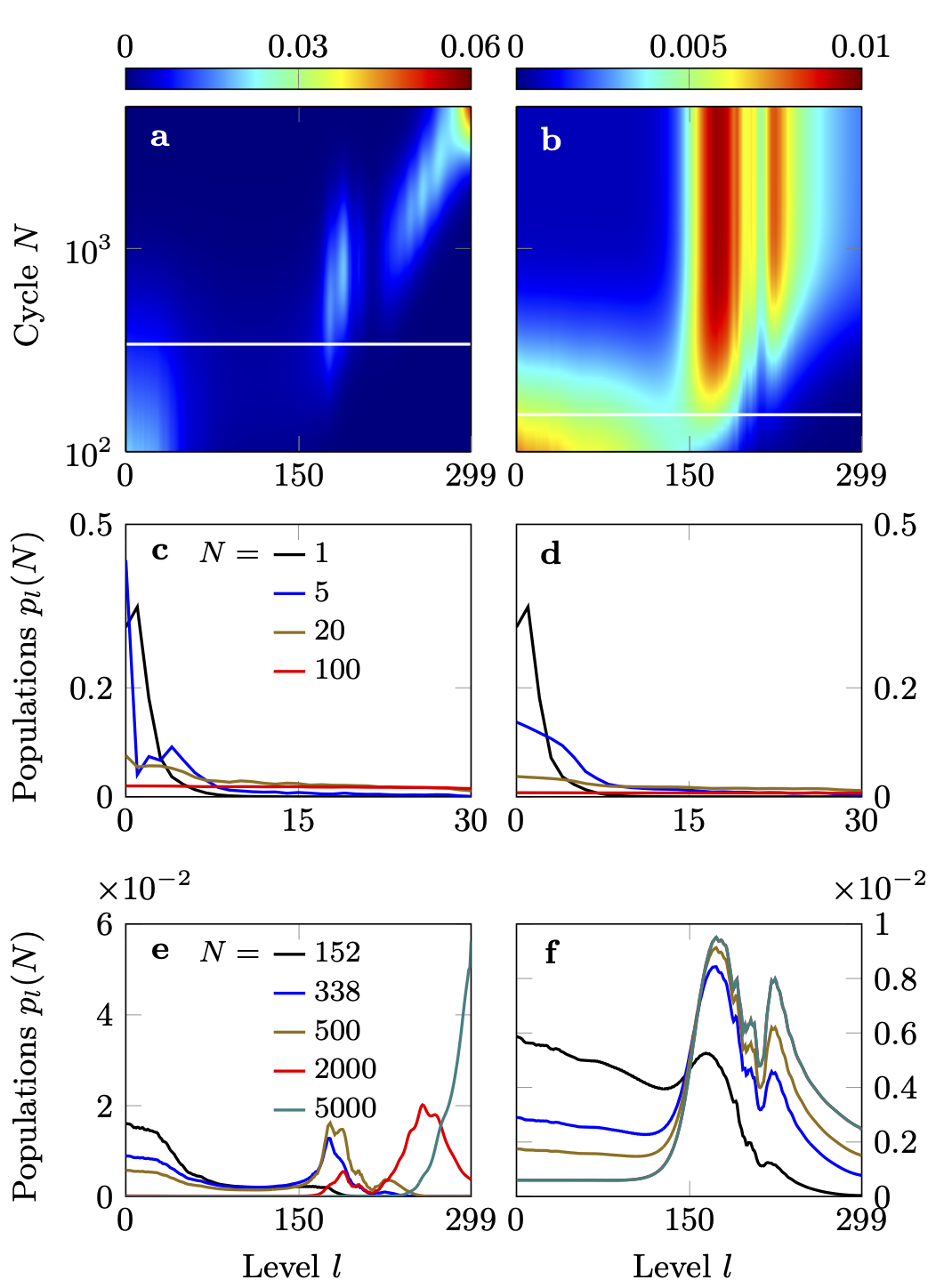}
    \caption{Populations $p_l(N) = \Tr \Pi_l \rho_N$ and $\tilde{p}_l(N) = \Tr \Pi_l \tilde{\rho}(N)$ of the $l$th battery state as functions of $l$ and $N$ for  unmeasured ({\bf a}) and measured  batteries ({\bf b}) and for a work-stroke duration $T_1 = 40 \pi \omega^{-1}$. All other parameters are chosen as in Fig.~\ref{fig:eng_WQ}. The left panels ({\bf c,e}) depict the populations $p_l$ as functions of the battery energy label $l$ for a selected set of cycle numbers and for an unmeasured battery. The panels ({\bf d,f}) present the according data for periodically measured batteries. The white lines in the panels ({\bf a,b}) indicate the respective critical cycle number $N^*$ which is 338 for the unmeasured and 152 for the measured battery. In both scenarios  a local maximum of the populations builds up after a sufficient number of cycles of the order of $N^*$. For the unmeasured battery this maximum grows and simultaneously moves with increasing $N$ to larger $l$ values until it reaches the highest possible energy battery state. In contrast, for the measured battery the population grows within an almost fixed region. While the initial population growth proceeds faster for the measured battery finally more energy is finally stored in the unmeasured battery.} 
    \label{fig:eng_pop}
\end{figure}
\begin{figure}[t!]
    \centering
    \includegraphics[width = \columnwidth]{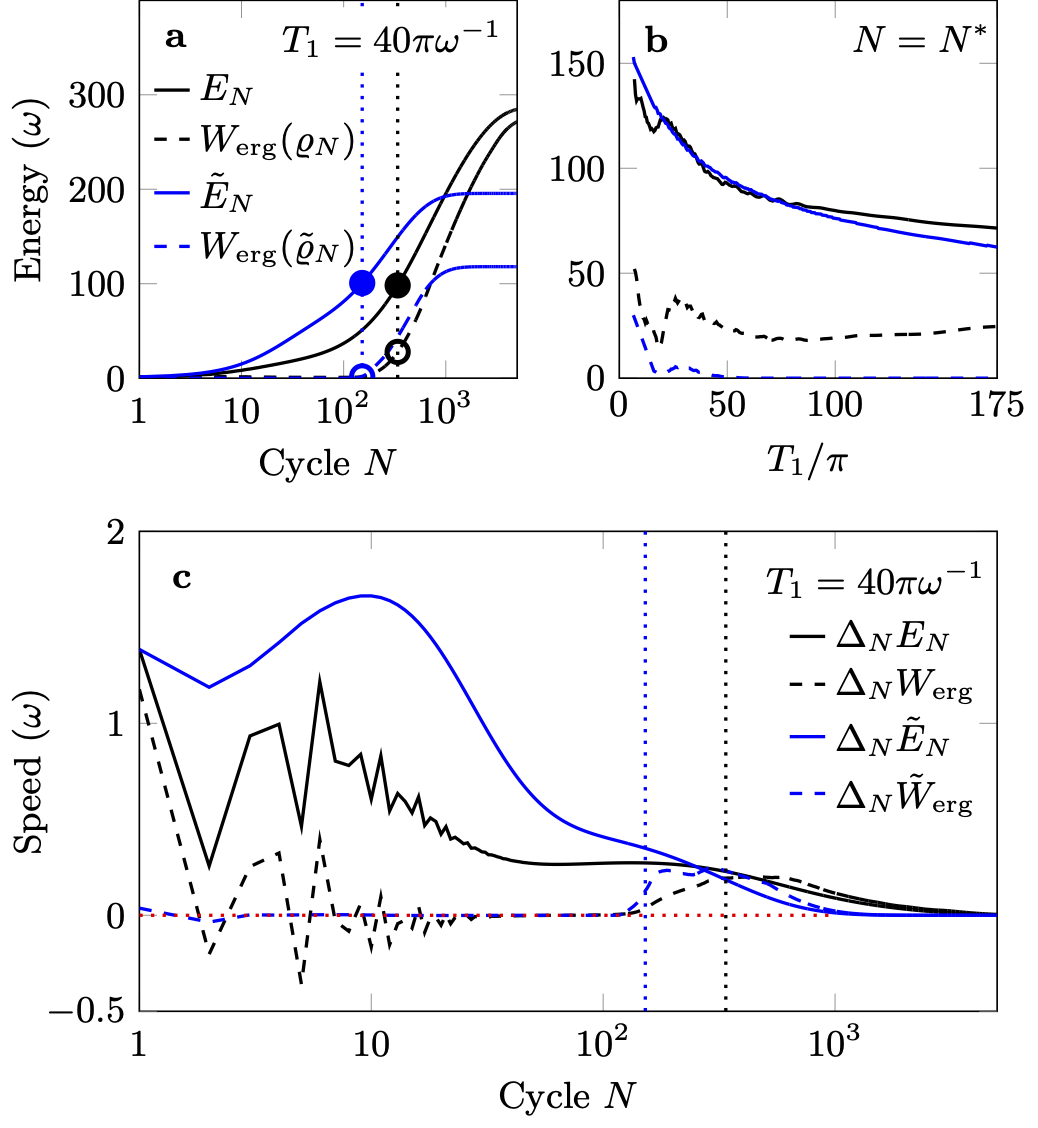}
    \caption{Average internal energy and ergotropy of an unmeasured and of a periodically measured QB are compared in panels {\bf a} and {\bf b}. In panels {\bf a}  and  {\bf c} the work-stroke duration is fixed at $T_1= 40\pi \omega^{-1}$ while the cycle number $N$ varies; the black and blue vertical dotted lines mark in the panels {\bf a} and {\bf c} the critical cycle number $N^*$ above which the measured and the unmeasured engines, respectively, fail to work for the given $T_1$. The corresponding battery energies and ergotropies are marked by full and open circles, respectively, in blue for the measured and in black for the unmeasured battery. In panel {\bf b} the average energy and the ergotropy are displayed as functions of $T_1$ for the cycle number $N=N^*(T_1)$, i.e., along the solid white curve in Fig.~\ref{fig:eng_WQ}. Panel {\bf c} depicts the charging speed, i.e., the change of the average energy and ergotropy per cycle as a function of the cycle number $N$ at $T_1 = 40\pi \omega^{-1}$. \jt{Note that the charging speed is proportional to the cycle averaged charging power.} The red dotted line indicates  zero speed. At negative speeds the battery is discharging. All the other parameters are the same as in Fig.~\ref{fig:eng_WQ}.}
    \label{fig:eng_etrans}
\end{figure}
The development of the population inversion is illustrated in Fig.~\ref{fig:eng_pop} in more detail. The upper row displays the populations of the considered battery with $M=300$ energy states, $p_l = \Tr_\mathrm{B} \Pi_l \varrho$ and   $\tilde{p}_l = \Tr_\mathrm{B} \Pi_l \tilde{\varrho}$, evolving as functions of the number of cycles for a fixed value of the work stoke time, $T_1 = 40\pi \omega^{-1}$, for which the charging of the unmeasured battery  is fastest.  The two lower rows display the charging process for the same $T_1$ in terms of the population of the battery energy states as a function of the number of cycles. 
In an unmeasured battery, three subsequently appearing separate  peaks characterize the charging progress of the battery  -- one around the ground state originating from the initial state and becoming insignificant around five hundred cycles, the second one that is formed around two hundred cycles and vanishes after three thousand cycles, and the last one emerging at around five hundred cycles that moves towards higher energies with increasing cycle number, asymptotically leading to a fully charged battery with perfect population inversion. 
In contrast, for a periodically measured battery, the peaks overlap and a multi-modal distribution builds up in the upper half of the energy spectrum while the populations of the lower levels keep decreasing. The highest energy level also gains population but remains less populated than the peak region lying around two thirds of the maximal level. This feature is qualitatively independent of the choice of $T_1$.

In spite of the fact that the coherent part of the ergotropy of the unobserved battery is negligibly small compared to its incoherent contribution, the mere presence of coherence and its undisturbed transmission between subsequent cycles leads to  fundamental differences in the temporal behavior of  most engine-battery metrics  in comparison to a periodically measured battery as it becomes evident from Figs.~\ref{fig:eng_WQ}--\ref{fig:eng_pop}.  A further comparison of the influence of the presence or the absence of battery energy measurements, and, hence, of hindered or free propagation of coherence, is presented in Fig.~\ref{fig:eng_etrans}. In the panel {\bf a}  the average battery energy content is displayed as function of the number of cycles confirming that the machine with a periodically measured battery is charged faster but also that it fails earlier to work as an engine than the same machine with an unmeasured battery. Within the regime of a working engine, i.e., for cycle numbers $N<N^*$, the measured battery is charged faster reaching a maximal average energy $ E_{N^*} $ comparable to that for the unmeasured battery, see the blue and black solid circles in Fig.~\ref{fig:eng_etrans}. On the other hand, the maximal ergotropies $W_{\rm erg}(\varrho_{N^*})$ (open circles) at the border of the working engine regime for an unmeasured battery is about fifteen times larger than that of a measured battery.
At larger cycle numbers the unmeasured battery takes over with respect to the average energy and the ergotropy.

The Fig.~\ref{fig:eng_etrans}{\bf b} displays the dependence of the average battery energy and ergotropy along the white solid line of the first engine failure $N^*(T_1)$ as a function of the work stroke time $T_1$. There is only little difference between measured and unmeasured batteries for the average energy. On the other hand, the ergotropy of the measured battery soon becomes negligibly small while that of the unmeasured battery approaches a roughly constant value \jt{which is approximately half of the total stored energy.} In panel {\bf c}, the speed of the loading process is displayed as the difference of the average energy and of the ergotropy during a cycle, $\Delta_N E =E_N -E_{N-1}$ and $\Delta_N W_{\rm erg} = \erg - W_\mathrm{erg}(\varrho_{N-1})$. The speed of the average energy varies for the measured battery in a rather smooth way while it strongly fluctuates during the first cycles of an unmeasured battery. \jt{The speed divided by the duration of the respective cycle gives the charging power averaged over the cycle. Hence, for a machine running with constant cycle duration -- as we silently assume here -- charging power and speed are proportional to each other.}

  While the speed of the average energy is always positive, the speed of the ergotropy may also assume negative values, in particular for the unmeasured battery at early cycles. The vanishing of the speeds of the average energies and of the ergotropy  after sufficiently many cycles indicates the reaching of the asymptotic state. In this final state the external drive delivers energy via the machine solely to the cold and the hot heat baths without altering the battery energetics.

\begin{figure}[t!]
	\centering
	\includegraphics[width=\columnwidth]{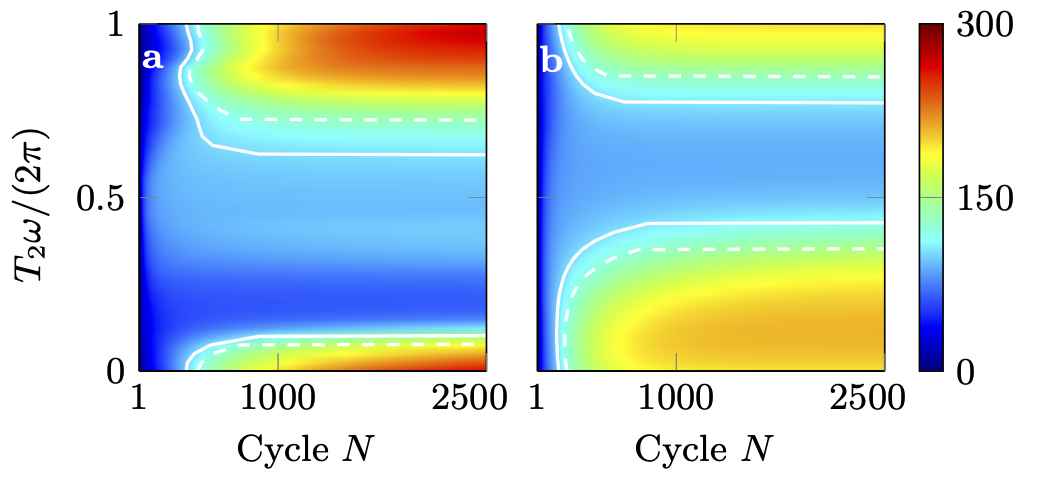}
	\caption{The average energy of the battery as a function of the thermalization time $T_2$ and the number of cycles for an unmeasured battery ({\bf a}) and a measured battery ({\bf b}). At any fixed sufficiently large number of cycles the average battery energy reaches the same maximal value at $T_2=0$ and $T_2=2 \pi /\omega$ presenting the upper and lower borders of both panels. \jt{These maxima are relatively flat such that close to the upper and lower borders still a large average energy of the battery is found even in the absence of perfect matching. This effect is more pronounced for the measured battery.} For other times $T_2$, the average battery energy may become substantially smaller. This effect is stronger for the unmeasured battery but is also present for the measured case. In the regions of suppressed battery energy the engine performs work on the external agent even at cycle numbers at which it long has stopped to work for thermalization times $T_2=0$ and $T_2=2 \pi /\omega$. As in the previous figures the full white line indicates the location at which the engine ceases to perform work on the external agent; when crossing the dashed line the working fluid delivers heat to both reservoirs. All parameters are the same as in Fig.~\ref{fig:eng_pop}.}
\label{FT2}
\end{figure}

So far all presented results refer to thermalization times $T_2$ that are integer multiples of the principal battery period $T_\omega =2 \pi/\omega$.  After these times the battery has exactly returned to its initial state. For thermalization times which strongly differ from integer multiples of $T_\omega$ we find a drastic reduction of the average battery energy as exemplified in Fig.~\ref{FT2}.
The same qualitative behavior can be observed for unmeasured and measured batteries whereby the effect is more pronounced for an unmeasured battery. In the regime of strong suppression the engine continues to work properly. Hence, a non-trivial time evolution of the battery during the thermalization strokes diminishes the interaction of the engine and the battery leading to lower average energy stored in the battery. \jt{Yet, there is a considerable level of tolerance with respect to the precise choice of the contact time $T_2$ as can be seen from Fig.~\ref{fig:eng_etrans}, in particular for a regularly measured battery.}             

\subsection{Initializing the Machine as a Refrigerator}\label{subsec:refrigerator}
In a wide range of parameters an Otto machine with a two-level working fluid may act as a refrigerator~\cite{Abah_2016, Han2020}, which pumps heat from the cold to the hot bath by means of work performed by the external drive on the machine. To operate accordingly, $\beta_h \epsilon_h> \beta_c \epsilon_c$ is required, meaning that the population of the excited state immediately after the completion of the hot heat stroke is smaller than that after the cold heat stroke. Furthermore, roughly speaking, the work strokes must not proceed too rapidly or too forcefully. For details, see the Append.~\ref{appendix:engine_only}. 

Similarly to the engine case (Sec.~\ref{subsec:engine}), the performance of a refrigerator is degraded by the addition of a load, such as a QB, during its work-strokes with the consequence that the machine ceases to work as a refrigerator beyond a critical number of cycles $N=N^*$ meaning that $\langle Q_{N<N^*}^c\rangle > 0$ and $\langle Q_{N>N^*}^c\rangle < 0$. 

\begin{figure}[t!]
	\centering
	\includegraphics[width=\columnwidth]{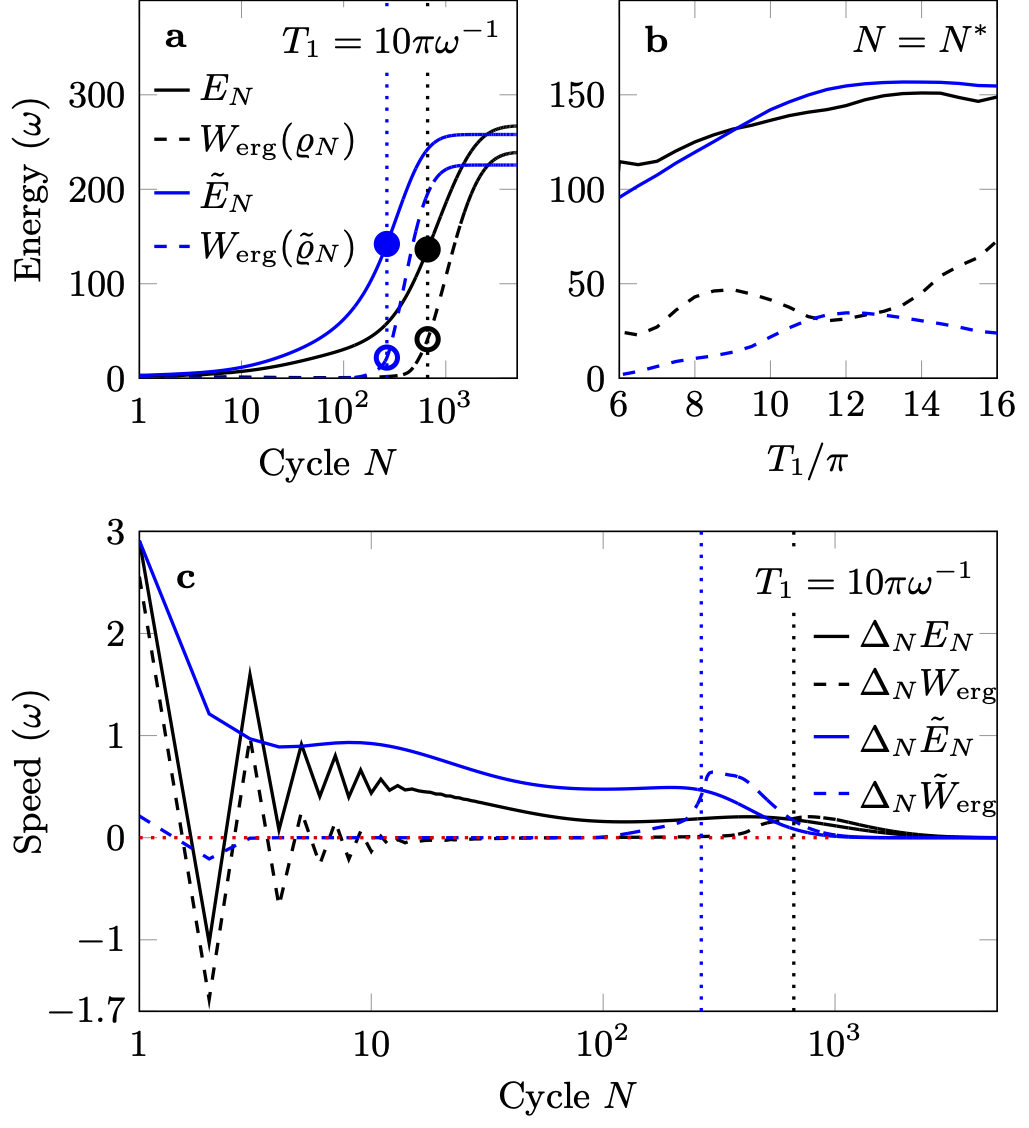}
	\caption{Replica of Fig.~\ref{fig:eng_etrans} with the exception that the machine is now initiated as a refrigerator. The average internal energy and ergotropy of the measured and unmeasured QB are plotted in panel {\bf a} as functions of the cycle number, whereas the same quantities are plotted in {\bf b} as  functions of the work-stroke duration when the refrigerator ceases to  extract heat from the cold reservoir. Panel {\bf c} depicts the charging speed for the energy stored and ergotropy of the QB. The vertical dotted lines in {\bf a} and {\bf c} indicate the respective critical cycle numbers $N^*$. The energy gap of the working fluid is modified to perform cooling as $\epsilon_c = \Delta = 10\omega$, $vT_1 = 300\omega$, and the rest of the parameters $g = \omega$, $\beta_h^{-1} = 200\omega$, and $\beta_c^{-1} = 20\omega$ remains the same as for the engine case, i.e., $\epsilon_h \simeq 300\omega$, $\alpha \simeq 5\cdot 10^{-15}$, $\beta_c\epsilon_c = 1/2$, $\beta_h\epsilon_h \simeq 3/2$.}
	\label{fig:ref_etrans}
\end{figure}

\begin{figure}[t!]
	\centering
	\includegraphics[width=\columnwidth]{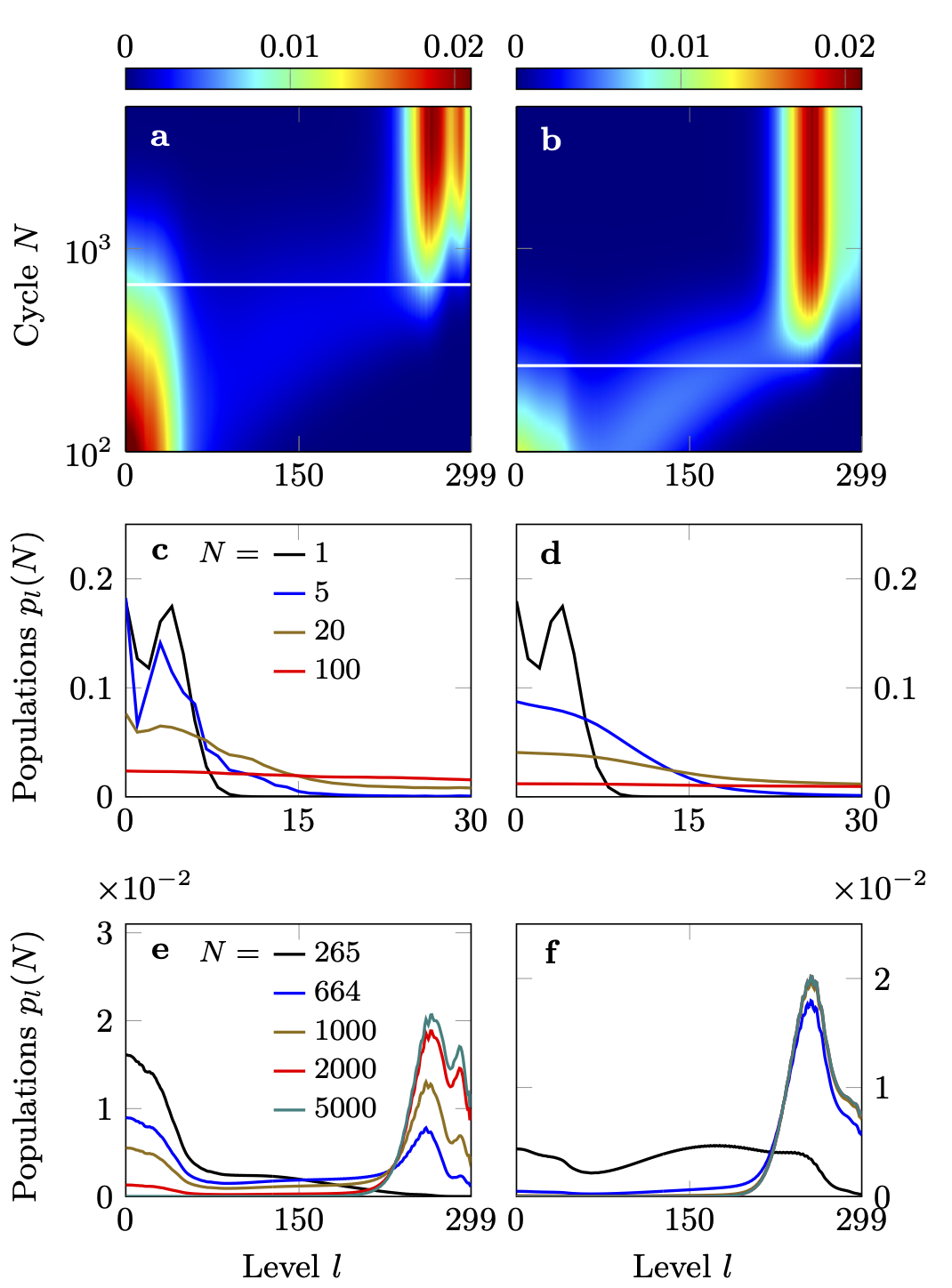}
	\caption{Replica of Fig.~\ref{fig:eng_pop} with the machine initialized as a refrigerator. Populations $p_l(N) = \Tr \Pi_l \rho_N$ and $\tilde{p}_l(N) = \Tr \Pi_l \tilde{\rho}(N)$ are calculated for the machine operating with a work-stroke duration $T_1 = 10 \pi \omega^{-1}$.  All other parameters are chosen as in Fig.~\ref{fig:ref_etrans}. The white lines indicate the respective critical cycle number $N^*$ which is 664 for the unmeasured and 265 for the measured battery.}       
	\label{fig:fridge_pop}
\end{figure}

We consider the same metrics for the refrigerator-battery system as for the engine-battery system in Sec.~\ref{subsec:engine}. As for an engine-charged battery, during a first epoch, the periodically measured refrigerator-charged battery is faster charged  than an unmeasured battery. Also after the machine has stopped to work as a refrigerator the battery is further charged. The measured battery reaches an asymptotic population of its energy states while the unmeasured battery is still charging until it also approaches an asymptotic state. Both the average energy and the ergotropy are comparable to the respective asymptotic values for a measured battery as can be seen in Fig.~\ref{fig:ref_etrans}{\bf a}. In contrast to the engine charged battery, the asymptotic energy populations are quite similar for measured and unmeasured refrigerator charged batteries as illustrated in Fig.~\ref{fig:fridge_pop}. Likewise, the dependence of the average energy on the duration of the work stroke at the respective critical number $N^*$ of cycles does not differ much for a measured and an unmeasured battery. Only the differences of the ergotropies are more pronounced as displayed in Fig.~\ref{fig:ref_etrans}{\bf b}. The charging speed of the average energy and the ergotropy for the unmeasured battery fluctuates strongly during the first cycles and even assumes negative values indicating that the battery is discharged temporarily, as evident from Fig.~\ref{fig:ref_etrans}{\bf c}. In contrast, the charging speed of the average energy of a measured battery, \jt{and with it the charging power,} vary gradually as  functions of the number of cycles with an overall decreasing tendency. The speed of the ergotropy is almost vanishing during the first charging episode, reaches a maximum around $N^*$, and later decays to its vanishing asymptotic value.    

\section{Conclusions}\label{sec:conclusion}
In this work, the charging of a quantum battery possessing a finite number of energy states by means of an Otto machine with a two-level working fluid was studied. Without the battery such a machine can function as an engine or a refrigerator depending on the proper choice of certain parameters. The charging of the battery takes place via an interaction of the machine and the battery during the work strokes. In order to prevent a leakage of the battery energy to the baths as well as a possible decoherence of the battery state, the interaction between the machine and the battery is turned off during both heat strokes. We found that in this way an Otto machine may charge a battery in either mode of operations.    

In order to monitor the energetic metabolism of the machine-battery system we determined on the side of the machine those averaged energies that are exchanged as heats with the cold and hot bath and the total average work that is performed during the work-strokes on the working fluid per cycle.   The monitoring of these contributions can be realized by means of projective energy measurements of the working fluid in the beginning and at the end of each work-stroke while the working fluid is isolated from the heat baths. Under the assumption of perfectly thermalizing heat strokes these measurements do not influence the dynamics of the machine because they are always only performed  at times at which the machine states are diagonal with respect to the machine's energy basis. The amount of energy stored in the battery, which initially is prepared in its ground state, is measured either after the completion of each cycle or only once at the end after a prescribed number of cycles. Apart from the increase of the average battery energy, the ergotropy  as the maximal average energy difference that can be achieved  by any unitary transformation of the battery was determined and separated into its coherent and incoherent parts.                    

While the periodic measurements of the battery completely suppress any coherence at the end of each cycle, in an unmeasured battery a coherence building up during the loading process can be freely transferred from one cycle to the next one. Yet, the coherent contribution to the ergotropy remains negligibly small compared to the incoherent part. Nevertheless, the charging process is essentially influenced by the kind of monitoring. For both modes of machine operations, namely as an engine or a refrigerator, the charging is accelerated by periodic battery energy measurements, hence, reaching an asymptotic battery state after a fewer number of cycles  than without measurements. For a refrigerator these asymptotic states are characterized by quite similar populations of the few highest energy levels for periodically measured as well as for unmeasured batteries. In contrast,  only  an unmeasured engine loaded battery eventually  becomes virtually fully charged, while a measured battery asymptotically may only be partially charged.  

In both modes of operation, as an engine or a refrigerator, and for both monitoring schemes, the presence of the battery has qualitatively the same degrading influence on the performance of the machine; after a characteristic number of cycles the machine even starts failing to work according to its initial design: an engine no longer performs work and a refrigerator fails to extract heat from the cold reservoir. However, the battery continues being charged during the machine-failing period until finally an asymptotic battery state is reached.

In order to achieve an optimal charging of a battery with equidistant level spacing it is of utmost importance that the duration of the thermalization strokes during which the battery undergoes a unitary, free evolution approximately matches an integer multiple of the free battery principal period. For \jt{noticeably deviating,} non-commensurate thermalization times the free battery dynamics  effectively  diminishes the coupling to the machine setting in during the subsequent work stroke. As a consequence, the battery then 
may not be charged while at the same time the machine continues performing according to its initial design as an engine or a refrigerator.  Whether such optimal thermalization times can also be achieved for batteries with other than equidistant level spacing is an interesting question of future research.   

We conclude by emphasizing that the monitoring of the charging process of a quantum battery  may have a strong influence on this process and the finally achieved battery state. Of course there are many other monitoring strategies, be it via measurements taking place not after each but only after each $N$th cycle or not by projective but by generalized measurements. In any case, the adequate monitoring must be allowed for in the analysis, and, vice versa, the kind of monitoring has to be adapted to the practical requirements on the charging process such as speed or maximal capacity.

\begin{acknowledgments}
This research was supported by the Institute for Basic Science in South Korea (IBS-R024-Y2).
\end{acknowledgments}

\appendix
\section{Otto cycles operating with linear driving work-strokes}\label{appendix:engine_only}

In this appendix, we provide the analytic results for the average heat and work when the machine operates without the load of a QB. 
For the present two-level working fluid undergoing a variation of the level-distance, which is linear in time, one obtains with the analytic Landau-Zener results of Refs.~\cite{LZ_inter,Thingna17,Thingna_LZ} the following time-evolution operator  
\begin{eqnarray}
	U &=& \sqrt{1-\alpha}\left(e^{-i\phi}\lvert +_h\rangle\langle+_c\rvert + e^{i\phi}\lvert -_h\rangle\langle-_c\rvert\right)\nonumber\\ 
	&-& \sqrt{\alpha}\left(\lvert+_h\rangle\langle-_c\rvert - \lvert -_h\rangle\langle+_c\rvert\right),
\end{eqnarray}
that governs the machine dynamics during the compression-strokes. An expansion-stroke as the time-reversed process of the compression-stroke is then described by the unitary operator $\tilde{U} = C U^{\dagger} C$ with the complex conjugation operator $C$. Above, the states $|\pm_x\rangle$ ($x=c,h$) are the eigenstates of the Hamiltonian at the beginning and end of work-strokes, i.e.,
\begin{eqnarray}
	H_{\rm M}(0) &=& \epsilon_c \left(\lvert +_c\rangle\langle +_c\rvert - \lvert -_c\rangle\langle -_c\rvert\right) \equiv H_{\rm M}(2T_1), \nonumber \\
	H_{\rm M}(T_1) &=& \epsilon_h \left(\lvert +_h\rangle\langle +_h\rvert - \lvert -_h\rangle\langle -_h\rvert\right),
\end{eqnarray}
with $\epsilon_c = \Delta$ and $\epsilon_h = \sqrt{\Delta^2+v^2T_1^2}$ using Eq.~(\ref{eq:H_engine}) and $\xi(t) = vt$. The parameters $\alpha$ and $\phi$ can be expressed as $\alpha = \exp[-2\pi\delta]$ and $\phi = \pi/4 - \delta(\log\delta-1) - \mathrm{arg}\Gamma(1-i\delta)$ with $\delta = \Delta^2/2v$ and the Gamma-function $\Gamma(z)$. The transition probability  $\alpha \in [0,1]$ vanishes for quasistatic process and becomes unity for a swap.

For the heat-strokes, we assume perfect thermalization into the canonical Gibbs states as defined by Eq.~(\ref{eq:Gibbs_state}). Starting with the working fluid in the cold state, the density matrices $\mu^T$ at the times $T$ indicated in Fig.~\ref{fig1:schematic} become
\begin{eqnarray}
	\mu^0 &=& \tau^c = \frac{e^{\beta_c\epsilon_c}}{Z_c}\lvert -_c\rangle\langle-_c\rangle +  \frac{e^{-\beta_c\epsilon_c}}{Z_c}\lvert +_c\rangle\langle+_c\rvert,\nonumber \\
	\mu^{T_1^-} &=& U\tau^c U^\dagger ,\nonumber \\
	\mu^{T_1^+} &=& \tau^h =  \frac{e^{\beta_h\epsilon_h}}{Z_h}\lvert -_h\rangle\langle-_h\rvert +  \frac{e^{-\beta_h\epsilon_h}}{Z_h}\lvert +_h\rangle\langle+_h\rvert,\nonumber \\
	\mu^{2T_1^-} &=& \tilde{U}\tau^h \tilde{U}^\dagger, \nonumber\\ 
\mu^{2T_1^+} &=& \mu^{0} ,\label{eq:after_WS2}
\end{eqnarray}
where $Z_x = 2 \cosh \beta_x \epsilon_x$, $x = h,c$. As it repeats periodically this single cycle describes the complete dynamics of an isolated machine.

The average heat and work per cycle follows from Eqs.~(\ref{eq:heat_def}) and~(\ref{eq:work_def}) as:
\begin{align}
	\langle W\rangle &= -\left[(1-2\alpha)\epsilon_h - \epsilon_c\right]\tanh\left(\beta_c\epsilon_c\right)\\
	&\  -\left[(1-2\alpha)\epsilon_c - \epsilon_h\right]\tanh\left(\beta_h\epsilon_h\right),\nonumber\\
	\langle Q^h\rangle &= \epsilon_h\left[(1-2\alpha)\tanh\left(\beta_c\epsilon_c\right) - \tanh\left(\beta_h\epsilon_h\right)\right],\\
	\langle Q^c\rangle &= \epsilon_c\left[(1-2\alpha)\tanh\left(\beta_h\epsilon_h\right) - \tanh\left(\beta_c\epsilon_c\right)\right].
\end{align}
Note that only the transition probability $\alpha$ but not the phase $\phi$ affects the thermodynamic quantities because any coherence built up in a work stroke is erased after each heat-strokes.

\begin{figure}
 \includegraphics[width=\columnwidth]{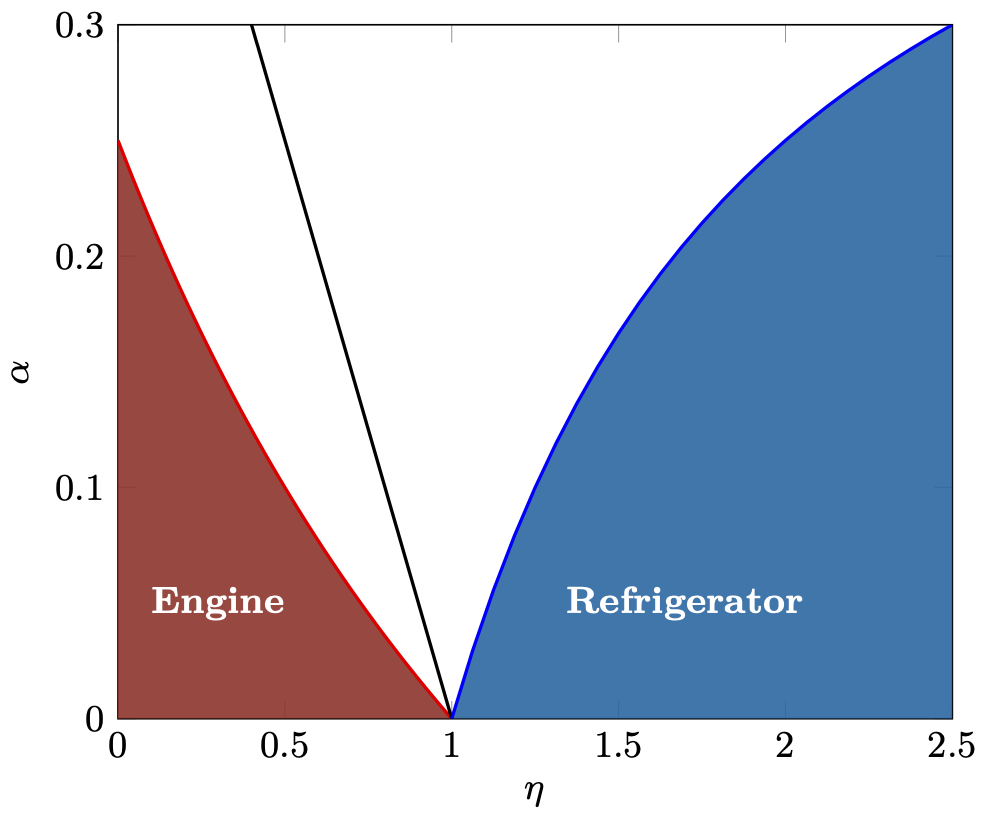}
\caption{Phase portrait of an isolated Otto engine on the $\alpha$-$\eta$ plane for the compression factor $x \equiv \epsilon_h/\epsilon_c =2$. The  machine acts as an engine in the red shaded region defined by 
the inequality~(\ref{W}). For parameters lying in the blue region representing the inequality~(\ref{Qc}) the machine acts as a refrigerator. The black line represents the limiting case of Eq.~(\ref{Qh}) that separates the two regimes of failing machines with either $\langle Q^h \rangle >0$ or $< 0$. For an equivalent phase portrait with a different parametrization see also Ref.~\cite{Solfanelli}.}           
\label{phs}
\end{figure}

For an engine the average work done on the working fluid must be negative ($\langle W \rangle <0$), the heat from the hot bath positive ($\langle Q^h \rangle >0$) and that from the cold bath negative  ($\langle Q^c \rangle <0$) implying for the engine parameters the conditions
\begin{align}
\alpha &> \frac{1}{2} \left ( 1 - \frac{x \eta +1}{\eta+x} \right ), \label{W}\\
\alpha &< \frac{1}{2} ( 1 - \eta), \label{Qh}\\
\alpha &> \frac{\eta -1}{2 \eta}, \label{Qc}
\end{align}
where $x = \epsilon_h/\epsilon_c >1$ and $\eta = \tanh \beta_h \epsilon_h / \tanh \beta_c \epsilon_c$. The condition $\alpha \geq 0$ leads together with (\ref{Qh}) to $\eta <1$ such that the third inequality (\ref{Qc}) is automatically satisfied. The allowed parameters for the machine as an engine are still restricted by the inequality (\ref{Qc}), see also Fig.~\ref{phs}.

For the machine to work as a refrigerator all inequality signs in (\ref{W} -\ref{Qc}) must be inverted. The accordingly inverted condition (\ref{Qc}) then implies with $\alpha >0$, $\eta >1$ and $\alpha < 1/2$. The other two inequalities are automatically satisfied, see Fig.~\ref{phs}.   

\begin{figure}[t!]
    \centering
    \includegraphics[width=\columnwidth]{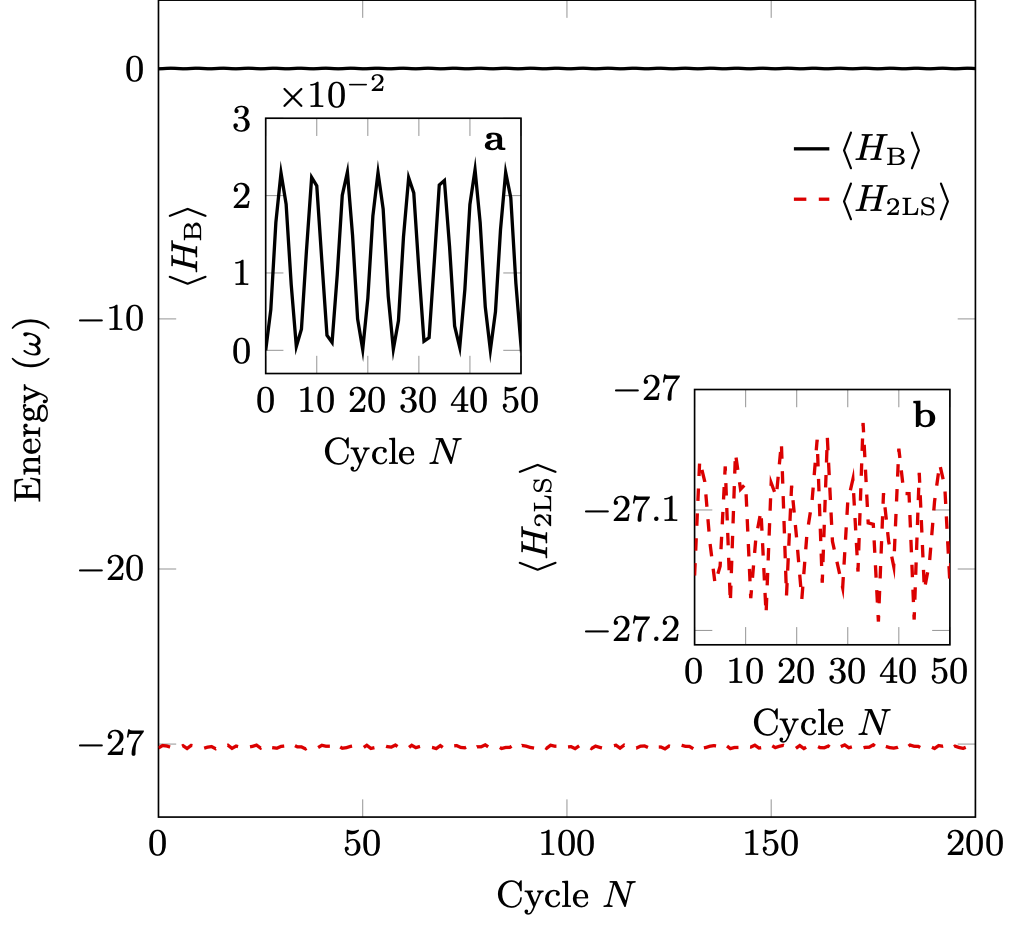}
    \caption{Average energy of the battery $\langle H_{\rm B}\rangle$ (black solid line) and of the two-level system $\langle H_{\rm 2LS}\rangle$ (red dashed line) as a function of the number of cycles $N$ for the model discussed in Append.~\ref{appendix:switch_onoff}. Small fluctuations are observed in the energies because the periodic variation of the interaction between the two-level system and the battery breaks the energy conservation. However, the energies do on average neither increase nor decrease with time $t$ indicating that the switching of the interaction has no influence on the charging of the battery. The parameters are chosen similarly as for the machine-battery setup of the main text with $\Delta = 30\omega$, $\epsilon = 200\omega$, $g = \omega$, and $M=30$. The switch-on period and switch-off period are set equal, i.e., $\tau_1 = \tau_2 = \omega^{-1}$. The two-level system is initiated in a state $\rho_{\rm 2LS}(0) = \exp [-\beta\Delta\sigma_x]/\Tr[\exp [-\beta\Delta\sigma_x]]$ with $\beta^{-1} = 20\omega$, similarly to the cold-bath temperature of the  Otto machine. Whereas, the battery is initiated in its ground state $\rho_{\rm B}(0) = |0\rangle\langle 0|$ with the battery and two-level system being decoupled initially.}
    \label{fig:appendix}
\end{figure}

\section{Work due to on-off switching}\label{appendix:switch_onoff}
In general, it is impossible to disentangle all the heat and work sources in the machine-battery composite to identify the contribution resulting from the on-off switching of the machine-battery interaction. We therefore study a simplified situation in which the on-off switching of an interaction between a two-level and an $M$-level system   provides the only potential source of energy. The Hamiltonian of the two-level system is of the form of Eq.~(\ref{eq:H_engine})  
\begin{eqnarray}
H_{\rm 2LS} = \Delta \sigma_x + \epsilon \sigma_z\:,
\end{eqnarray}
however, with a time-independent field $\epsilon$ in $z$-direction. The Hamiltonian of the  $M$-level system, $H_B$, is given by that of the battery, Eq.~(\ref{eq:H_battery}) and that of the time-dependent interaction by the Eq.~(\ref{eq:H_interaction}) with a coupling function of the form
 \begin{eqnarray}
g(t) &=& \begin{dcases}
        g & 0\leq t < \tau_1 \\
        0 & \tau_1 \leq t < \tau_2 \\
    \end{dcases}
\end{eqnarray}
which repeats periodically. 

Because there are neither work-strokes nor heat-strokes affecting the dynamics of this simplified system the switching of the interaction remains as a time-dependent contribution the sole potential source of an energy variation.

Figure~\ref{fig:appendix} displays the average energies of the battery and of the two-level system as a function of time over many periods. For the range of parameter values with $\epsilon, \Delta \gg g, \omega$, also always considered for the engine-battery system in the main part of this work, these average energies stay almost constant with negligible variations. Hence, the on-off interaction minimally influences the energies that remain constant over multiple cycles. If the interaction parameter $g$ becomes much larger than the battery level spacing $\omega$ the average energy of the two level system is greatly affected. Yet it apparently remains a bounded function of time as it is also the case for the average battery energy which oscillates with an amplitude scaling exponentially with $g$ for large $g$. Hence also in the strong coupling case the battery cannot be charged by solely switching the interaction. The simple model explored in this appendix strongly indicates that it is the action of the machine that charges the battery while the switching of the interaction has only minor consequences. \jt{Moreover, the results shown in Fig.~\ref{fig:appendix} do not depend on the duration of the switch-on switch-off periods. This strongly suggests that the presented  results regarding the machine and battery energies hold also in the presence of finite thermalization time as long as the system thermalizes in the heat strokes.}

\jt{
\section{Energetic cost of measuring the battery}\label{appendix:energetics}
According to the von-Neumann measurement scheme~\cite{vNeumann,Thingna2020} a pointer with a constant Hamiltonian, say $H_{\rm P}=0$, is brought in contact with the system in terms of the interaction Hamiltonian $H_{\rm BP}= \kappa A P$ for a time $\tau$ where $A$ is the system observable to be measured and $P$ the momentum operator conjugate to the pointer position. For any observable $A$ that commutes with the system-Hamiltonian $H_{\rm B}$, the interaction conserves the energy of the total system plus pointer. Additionally, since the Hamiltonian of the pointer is a constant, the energy of the system is thus naturally conserved. The initial density matrix of the system and the pointer factorizes whereby the latter is a minimum uncertainty state with vanishing position and momentum mean-values and position variance $\langle Q^2 \rangle$. In the position representation it hence takes the form $\sigma(x,y)=\exp[-(x^2 + y^2)/4\langle Q^2 \rangle]/\sqrt{2\pi\langle Q^2\rangle}$. Upon a position measurement of the pointer once the interaction has taken place one obtains a superposition of Gaussians centered around the eigenvalues of the considered observable $A$ all with the same variance $\sigma^2_A= \langle Q^2 \rangle/ \kappa^2 \tau^2$. In the limit of a vanishing variance, $\sigma^2_A $, i.e., for an infinitely precise pointer, the von-Neumann measurement scheme reduces to a projective measurement of the observable $A$, for details see \cite{Talkner16}.
 
In the setup of the Otto charging machine described in the main text, measurements of energies of the working fluid and the battery are performed at instants when the measured system is isolated from the working fluid and also from the heat baths. Because in such cases, with $A=H_{\rm B}$,  the interaction operator $H_{\rm BP}$ commutes with the full Hamiltonian $H_{\rm B} + H_{\rm BP} + H_{\rm P}$. Switching on and off the contact between the system to be measured and the pointer does not change the sum of their energies. Consequently, as the pointer has a constant energy, the energy of the measured system remains also unchanged. In other words there is no energy cost of a projective energy measurement on an isolated system.}

\bibliography{references.bib}  

\end{document}